\documentstyle[12pt, fleqn]{article}
\begin{document}

\title{\begin{flushright}
       {\normalsize 
        DOE/ER/40561--311--INT97--00--162 \\ }
        \end{flushright}
        \vspace{0.7 cm}
        \bf Importance of the direct knockout mechanism in
           relativistic calculations for
	   $\left(\gamma, p\right)$ reactions\thanks{
           This work was supported in part by the Natural Sciences and
           Engineering Research Council of Canada.} }

\author{ {\bf J.I. Johansson} \\
              Department of Physics, University of Manitoba \\
              Winnipeg, Manitoba, R3T 2N2 \\
              and \\
              Institute for Nuclear Theory, University of Washington \\
              Box 351550, Seattle, Washington, 98195-1550 \\
              and \\
         {\bf H.S. Sherif } \\
              Department of Physics, University of Alberta \\
              Edmonton,  Alberta, Canada T6G 2J1}

\date{\today}

\maketitle

\begin{abstract}

Results of relativistic calculations of the direct knockout (DKO)
mechanism for the photon induced removal of a proton from a target
nucleus over a wide range of energies and nuclei are presented.
Spectroscopic factors used in the calculations are fixed from
consistent analyses of the quasifree electron scattering process
$\left(e, e^{\prime}p\right)$.
The results indicate that within the uncertainties of the model, the
knockout contributions are generally close to the experimental data
for missing momenta below $\approx$ 500 MeV/$c$.
This is in disagreement with nonrelativistic analyses which often
find that the direct knockout contribution can be quite small
compared to the data and that meson exchange corrections can be
important.
The present study suggests that meson exchange current contributions
may not be as large when treated in a relativistic framework.
We also point out some difficulties we encountered in analyzing
the data for a $^{12}$C target at photon energies below 80 MeV.

\end{abstract}


\newpage

\section{Introduction}      \label{intro}

The reaction mechanism leading to the knockout of a single proton by
a real photon has been the subject of some debate recently.
Some nonrelativistic analyses \cite{IS94,Bo95,Mo95,St90}
suggest that the direct knockout (DKO) contribution may be very small
compared to the data.
Accordingly it was concluded that meson exchange current
(MEC) contributions must be the main mechanism responsible for the
observed cross sections.
Similar conclusions were reported earlier for the nonrelativistic
analyses carried out by Miller {\em et al.} \cite{Mi95} for ground
state transitions for the reaction
$^{16}O \left(\gamma, p\right) ^{15}$N at 60 and 72 MeV,
and by Ireland {\em et al.} \cite{Ir93} for $\left(\gamma, p\right)$
reactions on several nuclei for $E_{\gamma}$ near 60 MeV
(the same nuclei involved in the discussion reported in Ref. \cite{IS94}).
The above conclusions do not seem consistent with the findings by
Ryckebusch {\em et al.} \cite{Ry92}.
These authors find MEC effects to be relatively small for ground
state transitions.
The above statements illustrate the existing difficulty of arriving at
a consensus within the nonrelativistic framework as to the extent of
contributions from processes beyond simple direct knockout to ground
state transitions in $\left(\gamma, p\right) $ reactions.
Although the differing views stated above appear to be somewhat
dependent on the nuclear models used in the nonrelativistic
calculations, they are, however, symptomatic of our incomplete
understanding of the nature of the reaction mechanism for
photonuclear reactions.

These results are quite different from those of a recent
relativistic analysis of Johansson {\em et al.}  \cite{Jo96} who
find that, for an incident photon energy of 60 MeV, the DKO
contribution accounts for most of the observed data,
with no indication of any systematic sharp deviation from the data
at this energy.

In this paper we extend the analysis reported in Ref. \cite{Jo96},
for a photon energy of 60 MeV, to a much wider range of data. 
We consider several data sets for the $\left(\gamma, p\right)$ reaction
[and some data on the inverse reaction $\left(p, \gamma\right)$]
on a number of target nuclei and covering a range of photon energies
extending well into the $\Delta$-resonance region.
The spectroscopic factors and wave functions used in the calculations
are fixed at the values obtained from a parallel analysis of the
$\left(e, e^{\prime}p\right)$ reaction on the same target nuclei.
The objective of this study is to use this type of constrained 
analysis to gain some insight into the role of the DKO mechanism and
to see if a consistent description of the available data is possible.

Section \ref{rel_obs} outlines the relativistic calculations for the
direct knockout contribution to the $\left(\gamma, p\right)$ reaction.
Results of the calculations and details of the comparisons with data
are given in section \ref{disc}.
Our conclusions are given in section \ref{concl}.

\section{Relativistic Calculations}   \label{rel_obs}

The differential cross section due to the direct knockout
contribution to the $\left(\gamma, p\right)$
reaction has been given previously \cite{Jo96,LS88} but we 
provide it here again for ease of reference.
The relativistic expression for the differential cross
section leading to a specific final state of the residual
nucleus can be written as
\begin{eqnarray}
  \frac{d\sigma}{d \Omega_{p}} 
     &=& \frac{ \alpha }{ 4 \pi }  \frac{ M c^2 }{ \hbar c }
         \frac{ {\left| \mbox{\boldmath{$p$}}_{p} \right|} c }
              {E_{\gamma} } 
         \frac{ c } {v_{rel} }
         \frac{1}{R}
         { \frac{ {\cal S}_{J_i J_f} (J_B) }{ 2J_B + 1 } }
         \sum_{\mu M_B r }
         { \left| \epsilon_{r}^{\beta}
                  N^{\mu M_B}_{\beta} \right| }^2
    \label{gp_cross}
\end{eqnarray}
where $M_B$ and $\mu$ are the
spin projections of the bound and continuum protons.
We denote the four-momentum of the final proton $p_{p}$ and the four-momentum
of the incident photon as $q$.
The four-vector $\epsilon_{r}^{\beta}$ is the photon
polarization vector with two polarization states $r$, and summation is implied
over repeated greek indices.
The recoil factor $R$ is given in any frame by \cite {FM84}
\begin{eqnarray}
   R = 1 - \frac{ E_{p} } { E_{R} }
           \frac{1} {\left| \mbox{\boldmath{$p$}}_{p} \right|^2}
           \mbox{\boldmath{$p$}}_{p} \cdot
           \mbox{\boldmath{$p$}}_{R}     .
   \label{recoil_fact}
\end{eqnarray}
The four-momentum of the recoil nucleus is denoted by $p_{R}$.
The function $N_{\beta}^{\mu M_B}$ is
\begin{eqnarray}
  N_{\beta}^{\mu M_B} = \int d^3x \; 
        \Psi_{\mu}^\dagger \left( p_{p}, \mbox{\boldmath{$x$}} \right)
        \Gamma_{\beta}
        \Psi_{J_{B}, M_{B}} \left( \mbox{\boldmath{$x$}} \right)
        \exp \left( i \mbox{\boldmath{$q$}} \cdot
                      \mbox{\boldmath{$x$}} \right)     ,
   \label{nuc_mat}
\end{eqnarray}
where the wave functions of the continuum and bound nucleons,
denoted $\Psi_{\mu}$ and $\Psi_{J_{B}, M_{B}}$ respectively,
are solutions of the Dirac equation containing appropriate
potentials \cite{LS88}.
The $4 \times 4$ matrix $\Gamma_{\beta}$, operating on the
nucleon spinors, is given by
\begin{eqnarray}
   \Gamma_{\beta}
               = \gamma_{0} \left[ \gamma_{\beta}
                                   + \frac{i \kappa_{p}}
                                     {2M} \sigma_{\beta \nu}
                                     q^\nu \right]   .
\end{eqnarray}

The ingredients of the model are basically the same as used in an 
analysis of light to medium weight nuclei at 60 MeV \cite{Jo96}:
the bound state protons are described by solutions of a Dirac
equation containing the relativistic Hartree potentials of
Blunden and Iqbal \cite{BI87}, 
while the final state continuum proton is described by 
solutions of a Dirac equation containing complex 
phenomenological optical potentials obtained from fits to
proton elastic scattering data \cite{COPE}.
Given these potentials, the only parameters left to determine are
the spectroscopic factors.
For the light nuclei: $^{10}$B, $^{12}$C and $^{16}$O, we have
obtained the spectroscopic factors by fitting the results of our
$\left(e, e^{\prime}p\right)$ model \cite{Jo96,HJS95} to available
data. 
The $^{208}$Pb data are not suitable for analysis using this model
because of the lack of Coulomb distortions for the incident
and final electrons.
The spectroscopic factors used in this case are those of Udias
{\em et al.} \cite{Ud96} who have performed a relativistic analysis
of the $^{208}$Pb data.

In the following we show the results of our calculations compared
to the experimentally determined cross sections for several
nuclei covering a wide energy range.

\section{Results and Discussion}         \label{disc}

We have performed calculations for the $\left(\gamma, p\right)$
reaction on several target nuclei, over a wide range of energies.
These are compared to existing data in order to assess the extent to
which the direct knockout mechanism contributes to the observed
cross sections.
The ingredients of the calculations have all been determined
elsewhere and since there are no adjustments made, the results
can be considered as predictions of the model.

In the graphs to be discussed below there are curves corresponding
to several different calculations.
The description of the calculations represented by each of these 
curves is as follows:
\begin{enumerate}
\item  dashed curve --- energy- (E-) dependent parameterization
       of the Dirac optical potentials specific to a single nucleus
       \cite{COPE} while the bound state wave function is obtained
       through a Dirac-Hartree calculation \cite{BI87};
\item  dotted curve --- E-dependent parameterization of the
       Dirac optical potentials specific to a single nucleus and
       the binding potential has a Woods-Saxon form;
\item  solid curve --- energy- and mass- [(E+A)-] dependent
       parameterization of the Dirac optical potentials and
       the same Dirac-Hartree bound state wave function as 
       in curve 1 above;
\item  dot-dashed curve --- curve 3 divided by a
       factor of 2.0 to bring the model calculations
       close to the data.
\end{enumerate}
All the figures shown below use this designation of curves.
The first three are simply for calculations using a variety of
existing potential models in order to provide some feeling for the
sensitivity of the results to variations in these ingredients.
The dot-dashed curve is only relevant to graphs shown for the 
$^{12}$C target.

\subsection{$^{12}$C Target}

A considerable amount of data are available for this target.
We have made comparisons of our relativistic DKO model calculations
with these data, concentrating mainly on  ground state transitions.
These comparisons are shown in Figs. 1-3.
Data for four of the energies shown in Fig. 1:
$E_{\gamma}$ = 49.0, 58.4, 67.8 and 78.5 MeV were reported by
Springham {\it et al.} \cite{Sp90} and are obtained
using the tagged photon facility at Mainz.
The absolute magnitude of their cross sections was obtained by
normalizing the data at each energy to data taken by Mathews
{\it et al.} \cite{Ma76} for the $\left(\gamma, p_{0+1}\right)$
reaction, data which include both the ground and first excited 
state of the residual $^{11}$B nucleus.
The data of Aschenauer {\it et al.} \cite{As96_1},
at photon energies of $E_{\gamma}$ = 45.0 and 54.0 MeV
were obtained at the MAX-Lab at the University of Lund.
These data were normalized completely independently of any previous
experiment, and found to be consistent with existing data within
systematic errors.
The data at $E_{\gamma}$ = 73.5 MeV from Rauf \cite{Ra96}
were also obtained at the MAX-Lab.
These data were normalized to previous measurements including
those of Mathews {\it et al.} \cite{Ma76}.

The most obvious feature apparent in Fig. 1 is that the 
calculations tend to form a narrow band lying above the data.
Note that there is not much sensitivity to reasonable
variation of the ingredients of the model.
The light dot-dashed curve in all the figures shows the
solid curve divided by a factor of 2, and this brings the
curve close to the data in all cases.
The calculated curves have the correct shapes and the variation of
magnitude with incident photon energy also seems to be correct,
but the curves lie consistently above the data by a factor
of 2.

Figure 2 shows the differential cross section as a function of
photon energy for four different proton angles.
The experimental data are taken from Ruijter {\em et al.}
\cite{Ru96}; the experiments were also performed at the
MAX-Lab at the University of Lund.
Absolute normalization of these data was obtained
independent of any other experiment and the results were
found to be consistent with a large amount of other data.
Again the calculated curves lie above the data by close to 
a factor of 2.
We see that for photon energies above 40 MeV the energy and
angular dependence are quite well reproduced by our model,
but results lie above the data by a factor of 2.

In an attempt to compare to other data as well, we consider
two experiments in which the first excited state of $^{11}$B
at 2.12 MeV (1/2$^{-}$)
could not be resolved from the ground state.
The experiment of Mori {\it et al.} \cite{Mo95} was performed at
the Laboratory of Nuclear Science at Tohoku University.
The experiment of Harty {\it et al.} \cite{Ha95} was performed
at Mainz.
The differential cross sections of both experiments were normalized
without reference to any other experimental results and found
to be consistent with other data.
In order to compare to some data which do not contain the
first excited state, ground state data from other sources
were multiplied by a factor of 1.27 by both groups.
This factor is assumed to account  for population of the 
first excited state in $^{11}$B being $\sim$27\% as probable as
population of the ground state over the range of energies
considered in both experiments.
We adopt this factor in what follows.

In the left-hand column of Fig. 3 we show the measured energy
distributions of Mori {\it et al.} \cite{Mo95} as the circular
data points.
The triangular points are taken from the data of 
Harty {\it et al.} \cite{Ha95}.
The angles at which these data were taken do not coincide with the
angles of the experiment by Mori {\it et al.}.
For this reason the triangles shown at
66.0$^{\circ}$ represent cross sections which have been averaged
for proton angles of 63.3$^{\circ}$ and 68.4$^{\circ}$,
while on the graph labeled 91.1$^{\circ}$
we show data averaged for 88.5$^{\circ}$ and 93.5$^{\circ}$.
Note that the 88 MeV data point of Harty {\it et al.}
lies almost on top of the 87.8 MeV data point of Mori {\it et al.}
showing the consistency between the two data sets.
The curves are calculated assuming that the recoil nucleus is
left in its ground state, and then multiplied by 1.27.
The curves again lie above the data by a factor of 2
at low photon energies, but at the higher energies of the
Mainz experiment the calculations seem to move closer to the data.

The right-hand column of Fig. 3 shows the angular distributions
obtained by Harty {\it et al.} \cite{Ha95} compared to calculations
as discussed in the previous paragraph.
As the energy of the incident photons increases, the
calculations seem to move from lying above the data by
about a factor of 2, to falling within the error bars for the
data at the highest energy.
Of course, because of the large error bars at larger proton
angles the trend is not definitive, but it is suggestive.

The picture that is emerging for the status of the comparisons for
the $^{12}$C case can be further clarified by looking at the data
available for $\left(p, \gamma\right)$ reactions on this nucleus as
well as those leading to its formation as a residual nucleus.
The data are those obtained recently by Bright {\em et al.}
\cite{Br96} at Uppsala. 
Figure 4 shows comparisons to the data at proton energies of 98 and
176 MeV for ground state transitions to $^{12}$C and $^{13}$N
residual nuclei.
The former reaction is the inverse of the $\left(\gamma, p\right)$
reactions discussed above.
The comparisons for the $^{12}$C residual nucleus are shown on the
right hand side of the figure.
Using the same wave functions and spectroscopic factors as in Fig. 1,
we find that the calculations for $T_{p}$ = 98 MeV lie slightly above
the data at all angles except for the last point at
$\theta_{p} = $140$^{\circ}$.
At 176 MeV the calculations are  closer to the data except for the
large angles.
Thus the data for the inverse reaction confirm the behavior alluded
to above; at lower energies the calculations seem to overestimate
the cross sections.

The comparisons on the left-hand side of Fig. 4 present a somewhat
different picture. 
Using the maximum value for the spectroscopic factor, the relativistic
calculations for radiative capture on $^{12}$C are close to or below
the data.
With a more realistic value of the spectroscopic factor the calculations
will be further reduced in magnitude.
This situation is in clear contrast to the cases discussed above. 
It should be noted, however, that with a spectroscopic factor in the range
0.5--1.0 (the maximum possible value is 1.0 in this case),
the contributions of the knockout mechanism to the reaction
are substantial in the region of lower missing momenta.

It is worthwhile to point out here that there are also unresolved difficulties
for the $^{12}$C target in the $\left(e, e^{\prime}p\right)$ reaction,
in addition to the difficulties discussed in the current work.
The data from NIKHEF \cite{St88} are for kinematics with a fixed final
proton kinetic energy of $T_{p}$ = 70 MeV and nonrelativistic 
calculations shown in that paper cannot reproduce the shape of the
distribution in missing momentum.
In particular when the calculations are scaled to fit the peak for
positive missing momenta, the calculations fall below the data for
negative missing momenta.
Our relativistic calculations show exactly this behaviour and
the spectroscopic factor that we have used in this work is obtained
by matching to the positive missing momentum peak.
A proposed solution to this problem was to adjust the ratio of
transverse to longitudinal response functions, and when this ratio
was adjusted to $\approx$1.3 \cite{St88} the shape of the missing
momentum distribution was reproduced.
This problem was considered further by van der Steenhoven \cite{St91}
who found no justification for this enhancement factor.
The newer data from Mainz on this nucleus, reported by Blomqvist
{\em et al.} \cite{Bl95} have a higher final proton kinetic energy of
$T_{p} \approx$ 90 MeV.
In this case, with the increase in normalization of the data by
a factor of 1.19 \cite{Fr96}, both nonrelativistic and our
relativistic calculations can describe the shape of the measured
missing momentum distribution using the spectroscopic factor as
obtained from matching to the positive missing momentum peak of
the NIKHEF data.
This behaviour is consistent with the current results for the
$\left(\gamma, p\right)$ reaction on this nucleus: at low final
proton energies model calculations differ from the data and as
the final proton energy increases the calculations move closer to
experimental results.

The data for proton knockout from the $1p_{3/2}$ orbital in $^{12}$C
seem to indicate that at low energies a simple shell model description
is not adequate to explain the data.
A more complete description might possibly involve inclusion of the
deformed nature of the ground state wave function through a 
configuration mixing picture.

\subsection{$^{10}$B Target}

Data for both the $\left(e, e^{\prime} p\right)$ and
$\left(\gamma, p\right)$  reactions has been obtained for a
$^{10}$B target by de Bever \cite{Be93} at two different energies.
The data for knockout of a $1p_{3/2}$ proton leading to the 
ground state of $^{9}$Be are shown in Fig. 5.
Spectroscopic factors were obtained by scaling the model
calculations to the $\left(e, e^{\prime} p\right)$ momentum
distribution (or reduced cross section) data for the 
$T_{p}$ = 70 MeV case.
The other $\left(e, e^{\prime} p\right)$ and
$\left(\gamma, p\right)$ curves were then calculated without
any adjustment of the parameters.
It should be noted that the optical potentials used here are
parameterized using proton elastic scattering data on targets from
$^{12}$C to $^{208}$Pb.
As a result of a lack of proton elastic scattering data on $^{10}$B, 
we simply extrapolate the (E+A)-dependent potentials for use with a target
lighter than $^{12}$C.
The $\left(\gamma, p\right)$ calculations for this nucleus are quite
sensitive to changes in the potentials used to generate the nuclear 
wave functions.
In spite of this, and the fact that $^{10}$B is not a closed shell
nucleus, the DKO clearly produces results in the neighbourhood of the
data.

\subsection{$^{16}$O Target}

Figure 6 shows the differential cross section as a function of
proton angle for knockout of a valence proton from
an $^{16}$O target, leading to the ground state of $^{15}$N.
The photon energy range is the largest available, with eight 
energies in the range 60 MeV $\leq E_{\gamma} \leq$ 361 MeV.
The data come from three sources:
Miller {\em et al.} \cite{Mi95} provide data points at 
energies of 60 an 72 MeV,
while data shown at 60, 80 and 100 MeV are from Findlay and Owens
\cite{FO77}.
The high energy data for photons in the range 196 MeV
$\leq E_{\gamma} \leq$ 361 MeV are from 
Adams {\em et al.} \cite{Ad88}.
At low energies the calculated curves are generally close to the data,
reproducing the magnitude and shapes quite well.
For the higher energy data of Adams {\em et al.}
the calculations tend to be close to the data points at small angles
while falling below the data as the proton angle increases.
This is the behavior one expects if meson exchange processes
are going to become important as the missing momentum increases.

In order to remove some of the kinematic dependence from these 
curves we have calculated a {\em reduced cross section} by dividing
the differential cross section of Eq. (\ref{gp_cross}) by a
kinematic factor \cite{BGP81,IS94}:
\begin{eqnarray}
      2 \pi^{2} \alpha
      \frac{ \left| \mbox{\boldmath{$p$}}_{p} \right| E_{p} }
               {E_{\gamma} }
     \frac{1}{M^{2}}
      \left[   \left| \mbox{\boldmath{$p$}}_{p} \right|^{2}
               \sin^{2} \left( \theta_{p} \right)
             + \frac{1}{2} \kappa_{p}^{2} E_{\gamma}^{2} 
              \right]   .
  \label{reduce}
\end{eqnarray}

Figure 7 shows the reduced cross section as a function of missing
momentum for all the experimental data shown in Fig. 6, as well
as additional data provided by Leitch {\em et al.}
\cite{Le85}.
The curves are generated using the same ingredients as the 
solid curves of Fig. 6 but restricted to the kinematic range
covered by the data.
An interesting observation here is that the model results are
close to the data for missing momentum less than about 500 MeV/$c$.
The vertical dotted line indicates the momentum of a free proton
with kinetic energy equal to the charged pion mass.
The calculations start to fall below the data in this kinematic
region, which seems to be a good indication that we are seeing the
need for inclusion of pion exchange diagrams to the 
reaction mechanism, and provides some idea of where 
these diagrams become important.

\subsection{$^{208}$Pb Target}

Figure 8 shows results for proton removal from $^{208}$Pb, leading
to two doublets and one resolved state in $^{207}$Tl, for two
relatively low photon energies: 45 MeV and 54 MeV.
The data are from Bobeldijk {\em et al.} \cite{Bo95}.
These authors performed a nonrelativistic distorted-wave impulse
approximation (DWIA) analysis of the data
and found that the DKO contribution tends to lie up to a factor of 10 
below the data. Revised recent analyses \cite{As96_1,As96_2} indicate 
that this factor may have been unrealistic. 
Our present analysis, on the other hand, shows that the relativistic 
calculations do come close to predicting the correct magnitudes of the 
observed cross sections. Within the parameter uncertainties, it is 
evident that the DKO mechanism is the leading contributor to
the reaction at these energies.

\section{Conclusions}                   \label{concl}

In this paper we have presented relativistic calculations for
the $\left(\gamma, p\right)$ reaction and its inverse for a number
of target nuclei.
The results for the light targets cover a wide energy range,
while the results for the lead target are at low energy but
for a variety of final nuclear states.
The analysis was done in a consistent manner with no free parameters.
In all cases but one, $^{12}$C$\left(p, \gamma\right)$,
the spectroscopic factor is obtained from a parallel analysis of the
corresponding $\left(e, e^\prime p\right)$ data.

In cases of transitions with simple nuclear structure, relativistic
calculations indicate that the DKO mechanism is the main contributor
to the cross section for lower missing momenta.
For larger missing momenta one finds clear deviations indicating an
increased role for higher order processes such as 
meson exchange and $\Delta$--isobar contributions.

Nonrelativistic analyses often indicate that the contributions
from the DKO mechanism are small and that meson exchange
effects are sometimes dominant even at lower energies.
In contrast, the present relativistic analysis suggests
substantial contributions from the DKO mechanism to the cross sections 
over a wide range of energies.
The analysis also points out that meson exchange effects are
required at higher missing momenta.

In the course of this analysis we have found that in the case of 
the $^{12}$C target for photon
energies below 80 MeV, the relativistic calculations appear to
overestimate the cross section data by close to a factor of 2.
This situation is puzzling and may indicate either some complications
due to the structure of the $^{12}$C nucleus itself or to some
subtleties in the combined analysis of $\left(e, e^\prime p\right)$
and $\left(\gamma, p\right)$ for this target.
It must also be noted that these difficulties do not occur
for the spherical nuclei $^{16}$O or $^{208}$Pb, which are also
considered in the present work.
It is our feeling that the differences between theory and 
experiment at the lower proton energies for the $^{12}$C target
reflect the need for a proper description of the structure of
this nucleus to include the intrinsic ground state deformation.
The consistent approach based on a combined analysis of these two
reactions \cite{FO77_ii,IS94,Jo96} leads, in our view,
to the conclusion that the $^{12}$C ground state cannot
be adequately described by simple single particle configurations.

In the case of transitions with simple structure (mainly single particle)
our calculations indicate that meson exchange effects will not be
important until one reaches missing momentum near 500 MeV/$c$.
With the effort to push $\left(e, e^\prime p\right)$ reactions
towards this region of missing momentum it would be interesting
to see how important MEC effects will turn out to be in the
relativistic model.
Van der Sluys {\em et al.} \cite{Sl96} have considered this question
in a nonrelativistic random phase approximation (RPA) framework and 
found large contributions
from MEC's for larger missing momenta.

One point of interest is that the reactions discussed,
$\left(e, e^{\prime} p\right)$ and $\left(\gamma, p\right)$,
show different sensitivities to the description of the bound state.
This is probably due to the different range of missing
momenta sampled by the two reactions.
The $\left(e, e^{\prime} p\right)$ reaction has been
primarily concerned with low missing momenta where the bound wave
function is constrained by properties such as binding energy and
rms radius.
The bound state wave functions that we use show little difference in
momentum space for small momenta, and so it is not surprising that the
$\left(e, e^{\prime} p\right)$ results are very similar
in this region.
Differences between the bound state wave functions do arise, however,
for larger missing momenta in $\left(e, e^{\prime} p\right)$ and for the
inherently large missing momentum reaction $\left(\gamma, p\right)$.
This is not a surprise because this is where the nuclear wave function
is poorly constrained and the region where we see differences between
these bound state wave functions in momentum space.

A common criticism of the distorted-wave Born approximation (DWBA) 
approach, both relativistic and
nonrelativistic, is the lack of orthogonality of the bound and
continuum wave functions.
It is argued that this lack of orthogonality could invoke spurious
contributions to the cross sections.
The distorted continuum wave function is an approximation to the
many-body wave function of the nuclear system with appropriate boundary
conditions.
This approximation derives its support from the fact that the wave
function is constrained by proton-nucleus elastic scattering data.
Nonrelativistic RPA calculations do not suffer from this lack of
orthogonality, but the wave functions are not able to account for
the elastic scattering data.
A simple method for restoring orthogonality has been suggested by
Boffi {\em et al.} \cite{Bo84} and Ciofi Degli Atti {\em et al.}
\cite{At83}.
These authors find the orthogonality effects to be relevant mainly at
large angles.
It is likely that this feature will carry over into the
relativistic calculations and hence would not substantially change
the main characteristics of the present calculations.

The present results pose certain challenges for the relativistic approach.
If the DKO contributions are large then the data would suggest that the
MEC effects are suppressed in the relativistic models, at least at the
lower energies.
Relativistic models must explain this suppression and in the meantime
face the challenge of accounting for the observed relatively large
photoneutron cross sections.

Spin-dependent observables are likely to play an
important role in clarifying the reaction mechanisms. 
It should be noted here that the cross section angular distributions
for $\left(\gamma, p\right)$ reactions do not have much structure in most
cases.
The differences between competing models are then mainly
differences in magnitudes, and hence may be related to normalization
uncertainties in the models.
When we discuss spin-dependent observables these normalization
uncertainties cancel out and hence a better test of the model
is likely to result.

\section*{Acknowledgments}

One of us (H.S.S.) would like to thank the members of the Institute for
Nuclear Theory at the University of Washington for their
warm hospitality.
We would like to thank Derek Branford for providing us with the data
contained in Rauf's thesis \cite{Ra96} and for generously giving us
permission to show them before publication.
We are also grateful to L.J. de Bever and E.C. Aschenauer for allowing
us to use their data before publication.

\newpage

\section* {Figure Captions}

\noindent FIG. 1. Knockout of a $1p_{3/2}$ proton 
from a $^{12}$C target leading to the $^{11}$B
ground state.
Angular distributions for seven different photon energies
ranging from 45 to 78.5 MeV.
Hartree bound state wave functions are used \cite{BI87} 
and the proton optical potentials are from Ref. \cite{COPE}.
The data are from Refs. \cite{Sp90}, \cite{As96_1}
and \cite{Ra96}.
Curves as discussed in the text.

\vspace{2.5 mm}
\noindent FIG. 2. Knockout of a $1p_{3/2}$ proton 
from a $^{12}$C target leading to the $^{11}$B
ground state. Distributions in photon energy at four
fixed proton angles: $\theta_{p}$ = 30.0$^{\circ}$, 
60.0$^{\circ}$, 90.0$^{\circ}$ and 120.0$^{\circ}$.
The data are from Ref. \cite{Ru96}.
Curves as discussed in the text.

\vspace{2.5 mm}
\noindent FIG. 3. Knockout of a $1p_{3/2}$ proton 
from a $^{12}$C target leading to the $^{11}$B
ground and first excited states.
Left-hand column --- distributions in photon energy at four
fixed proton angles: $\theta_{p}$ = 30.6$^{\circ}$, 
45.8$^{\circ}$, 66.0$^{\circ}$ and 91.1$^{\circ}$.
Right-hand column --- angular distributions for five photon energies
The data are from Refs. \cite{Mo95} and \cite{Ha95}.
Curves as discussed in the text.

\vspace{2.5 mm}
\noindent FIG. 4.
Differential cross section as a function of 
photon angle for the $\left(p, \gamma\right)$ reaction on
$^{12}$C and $^{11}$B leading to the ground state.
Curves as discussed in the text.
The data are from Bright {\em et al.} \cite{Br96}.

\vspace{2.5 mm}
\noindent FIG. 5.
Single proton removal from the $^{10}$B target leading to the
ground state in $^{9}$Be:
upper figure -- the $\left(e, e^{\prime} p\right)$ reaction,
lower figure -- the $\left(\gamma, p\right)$ reaction.
Curves as discussed in the text.
The data are from de Bever \cite{Be93}.

\vspace{2.5 mm}
\noindent FIG. 6.
Differential cross section as a function of 
proton angle for the knockout of a $1p_{1/2}$ proton 
from a $^{16}$O target leading to the $^{15}$N ground state.
Curves as discussed in the text.
The data are from Refs. \cite{Mi95,FO77,Ad88}.

\vspace{2.5 mm}
\noindent FIG. 7. 
Reduced cross section as a function of 
missing momentum for the knockout of a $1p_{1/2}$ proton 
from a $^{16}$O target leading to the $^{15}$N ground state.
Data as in Fig. 6 and from \cite{Le85}.
Curves as discussed in the text.

\vspace{2.5 mm}
\noindent FIG. 8.
Differential cross section as a function of 
proton angle for the knockout of protons from different levels
in a $^{208}$Pb target.
Curves as discussed in the text.
The data are from Bobeldijk {\em et al.} \cite{Bo95}.

\begin{figure}
\begin{picture}(1100,400)(0,0)
\includegraphics{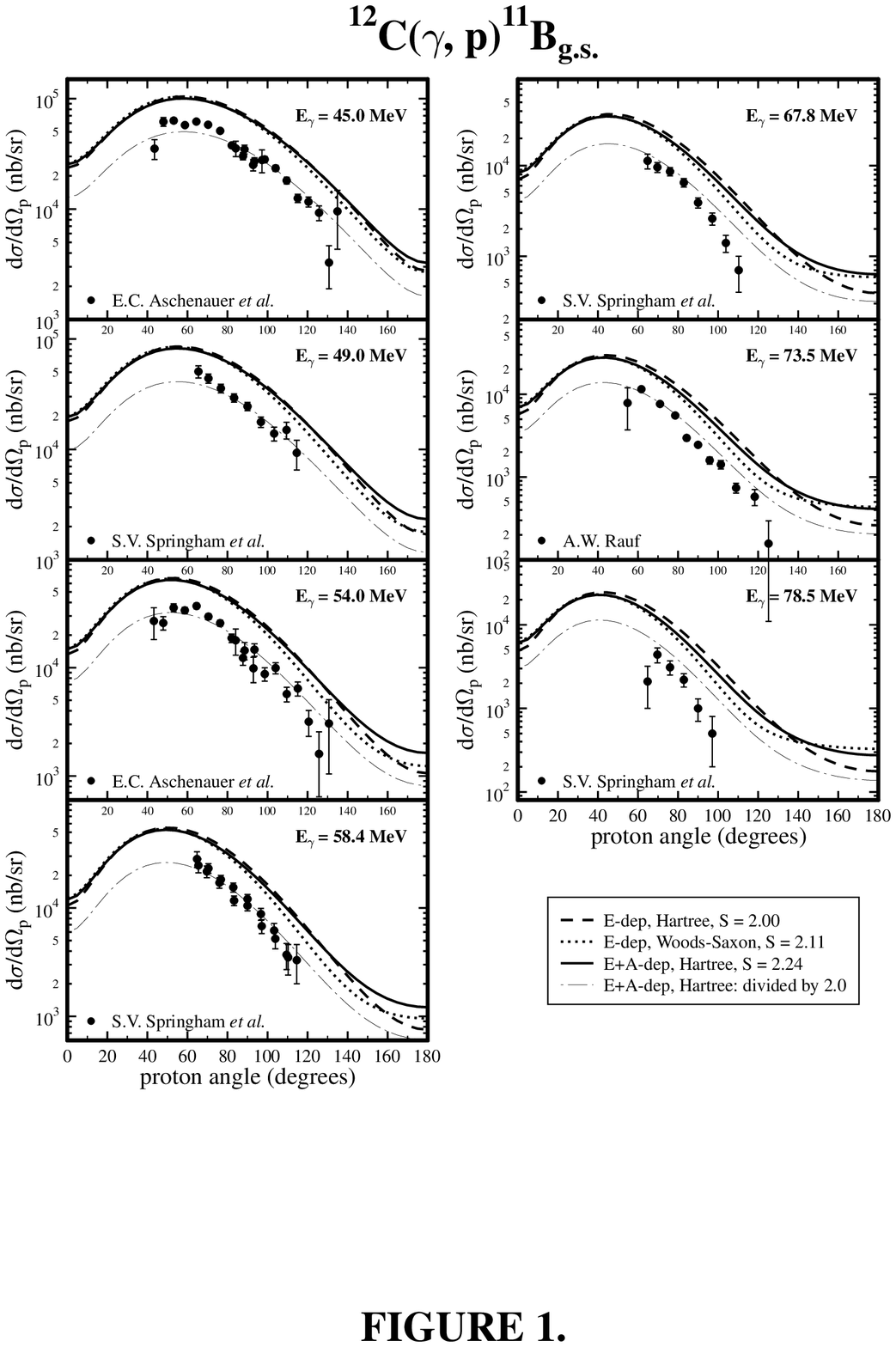}
\end{picture}
\end{figure}

\begin{figure}
\begin{picture}(1100,400)(0,0)
\includegraphics{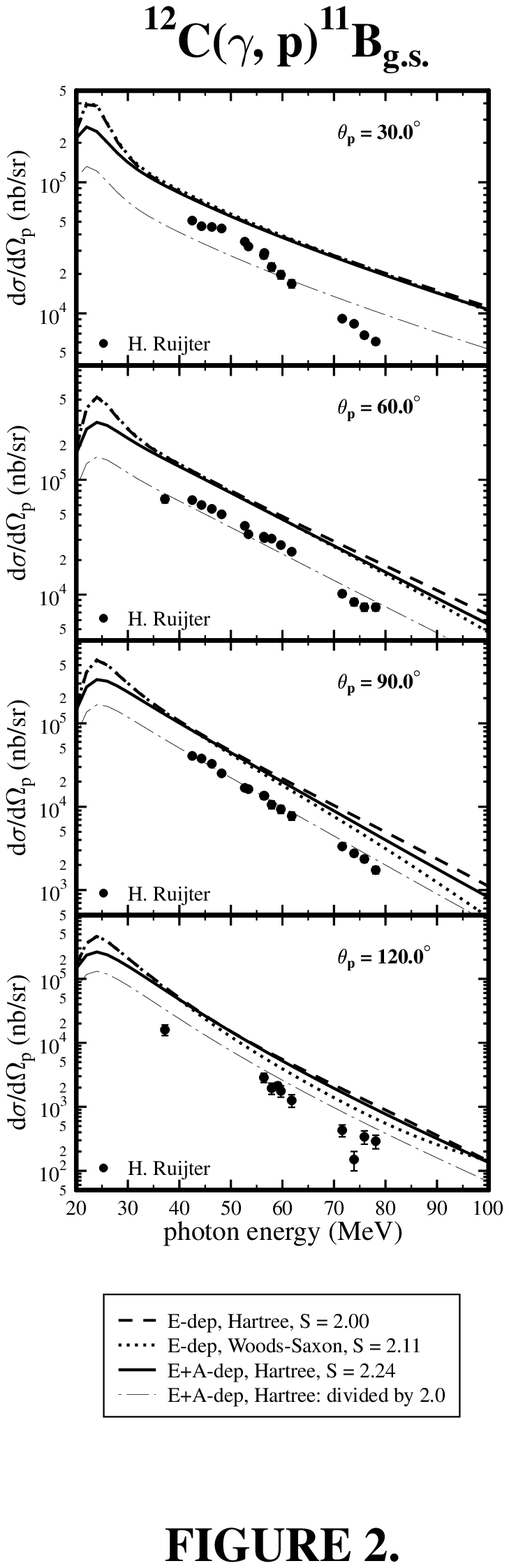}
\end{picture}
\end{figure}

\begin{figure}
\begin{picture}(1100,400)(0,0)
\includegraphics{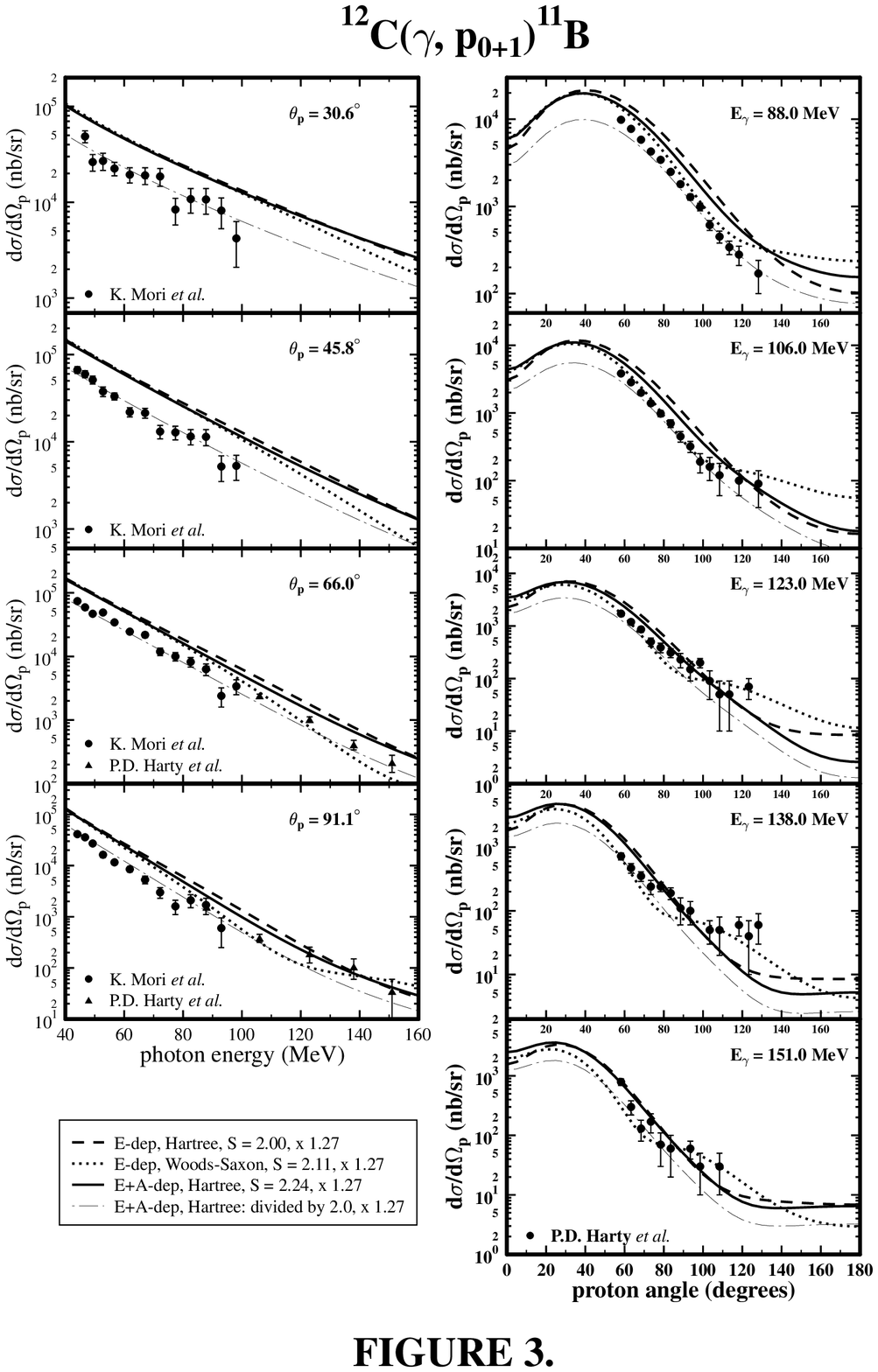}
\end{picture}
\end{figure}

\begin{figure}
\begin{picture}(1100,400)(0,0)
\includegraphics{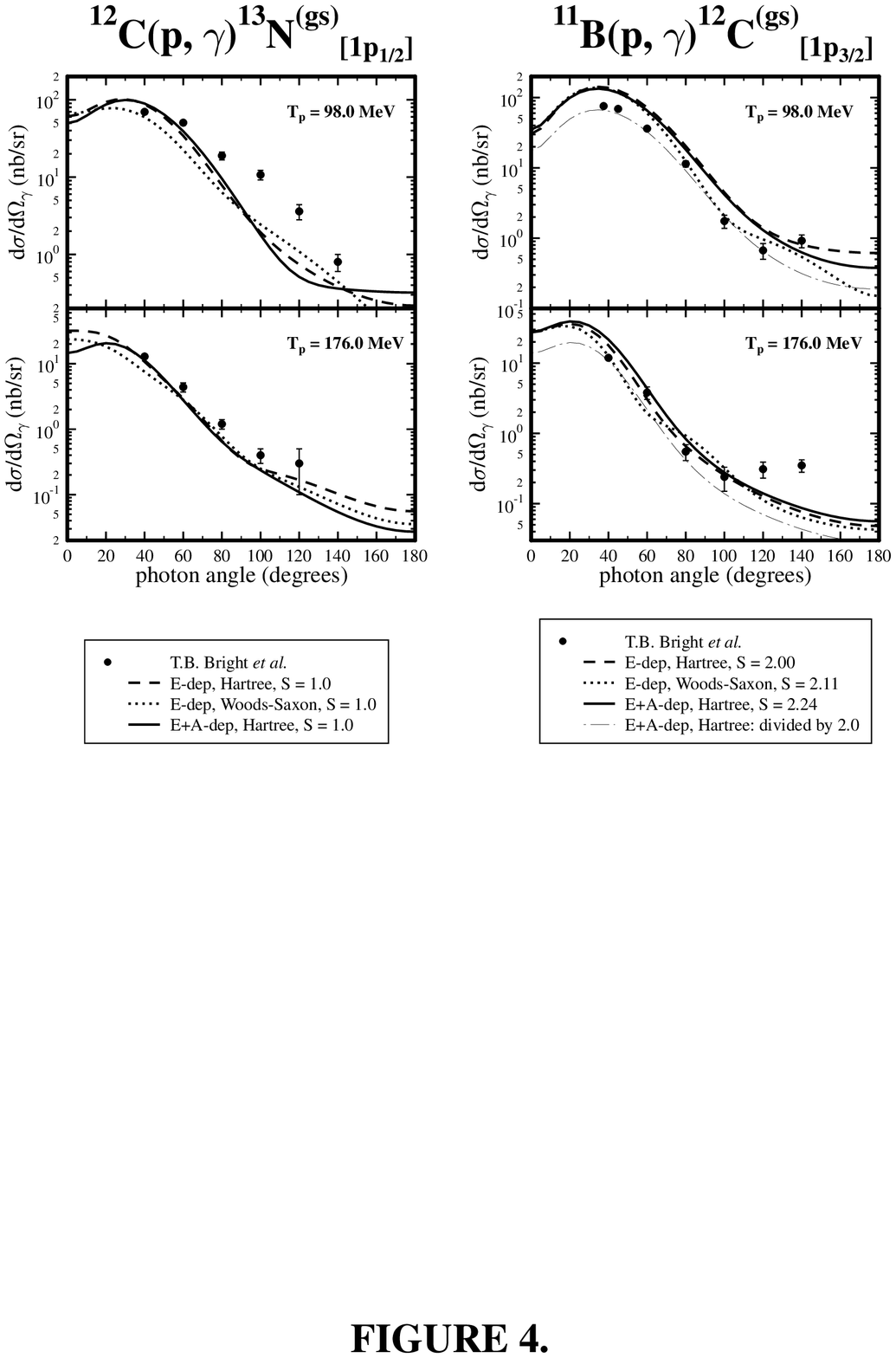}
\end{picture}
\end{figure}

\begin{figure}
\begin{picture}(1100,400)(0,0)
\includegraphics{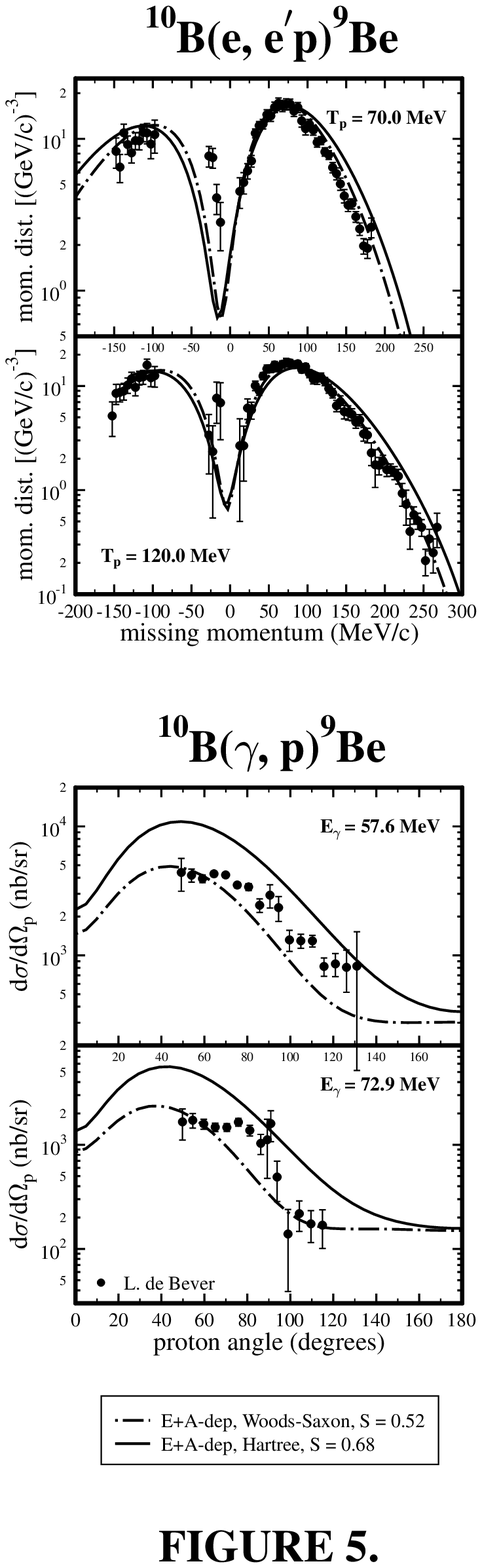}
\end{picture}
\end{figure}

\begin{figure}
\begin{picture}(1100,400)(0,0)
\includegraphics{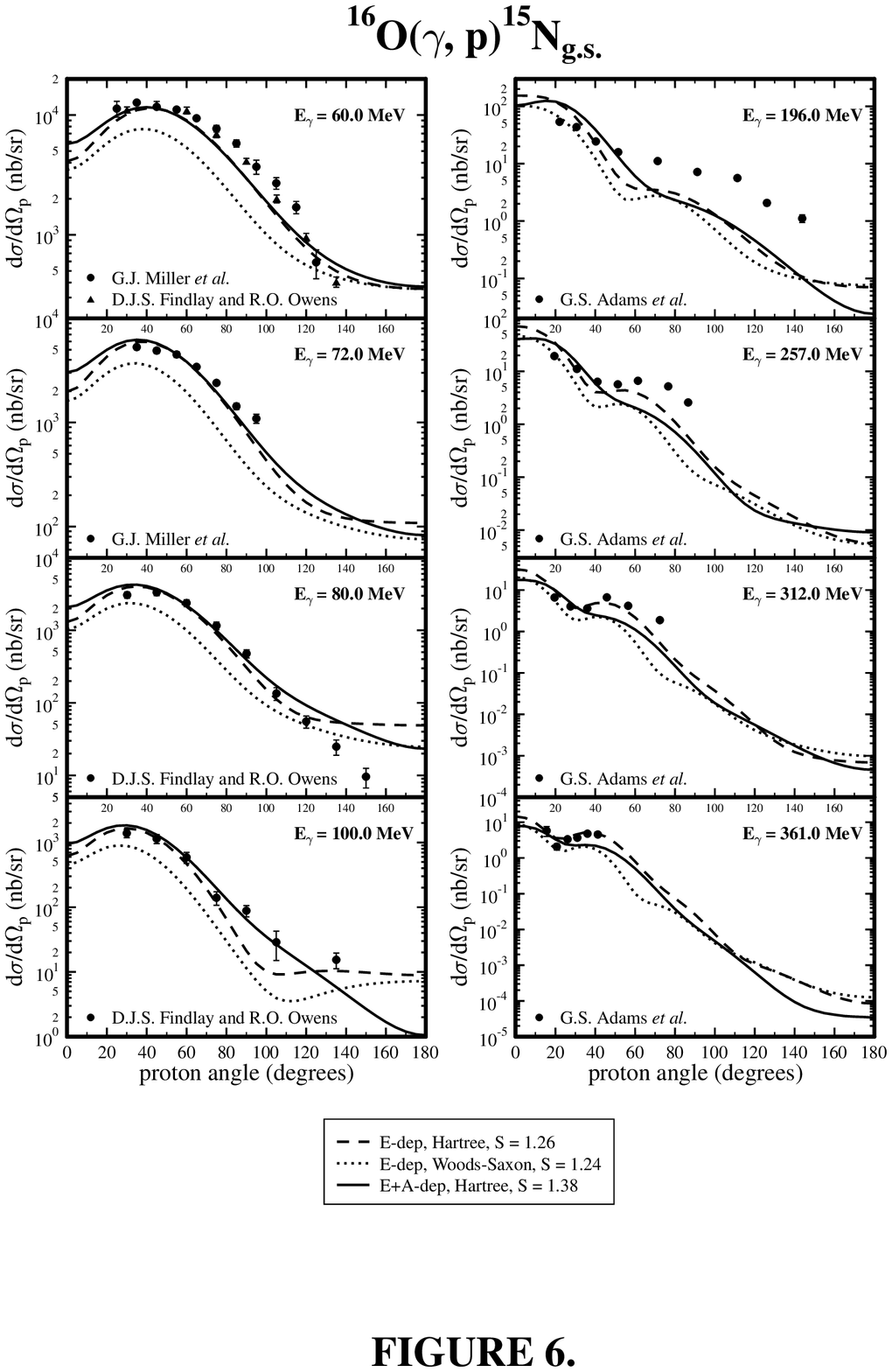}
\end{picture}
\end{figure}

\begin{figure}
\begin{picture}(1100,400)(0,0)
\includegraphics{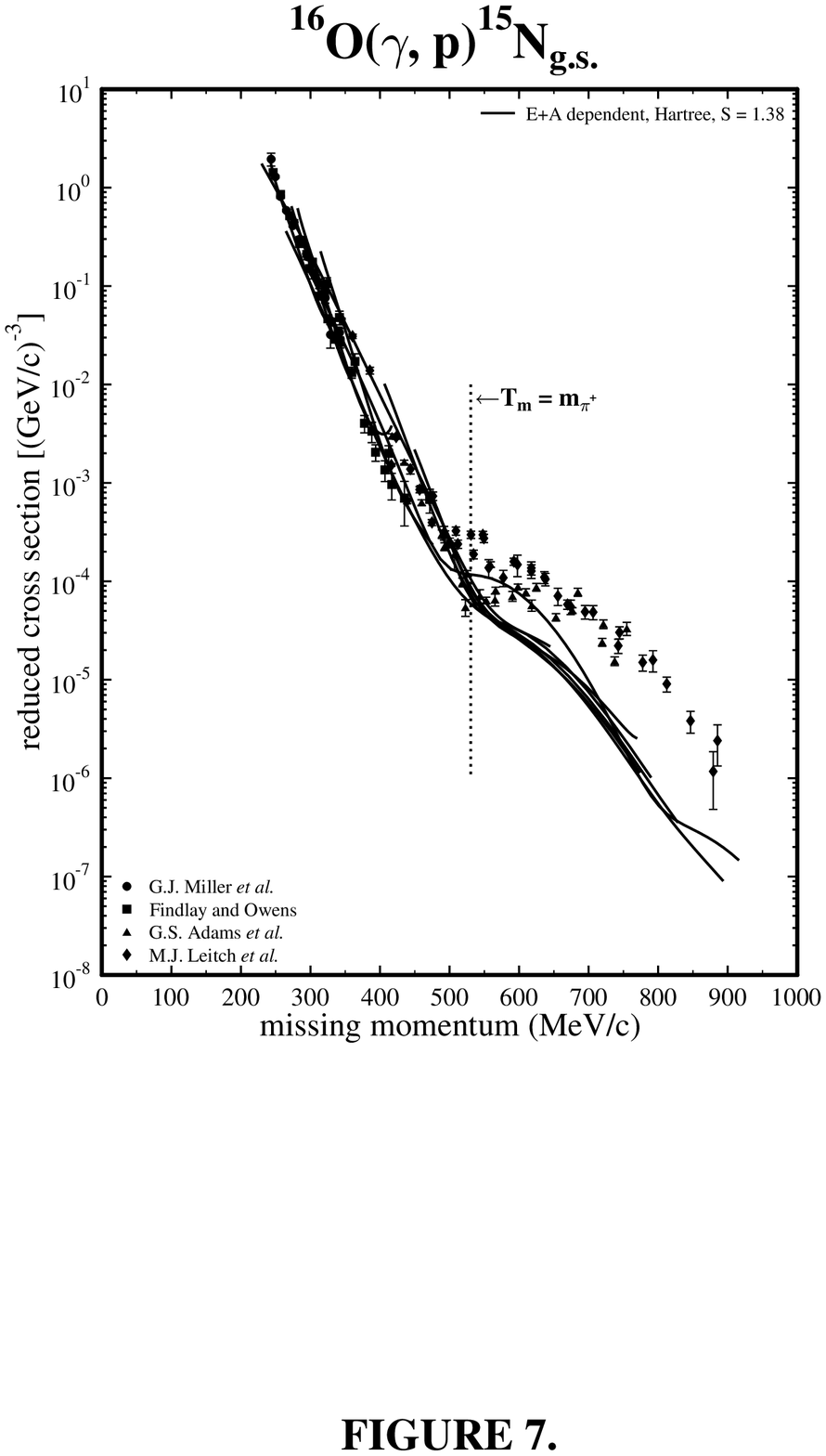}
\end{picture}
\end{figure}

\begin{figure}
\begin{picture}(1100,400)(0,0)
\includegraphics{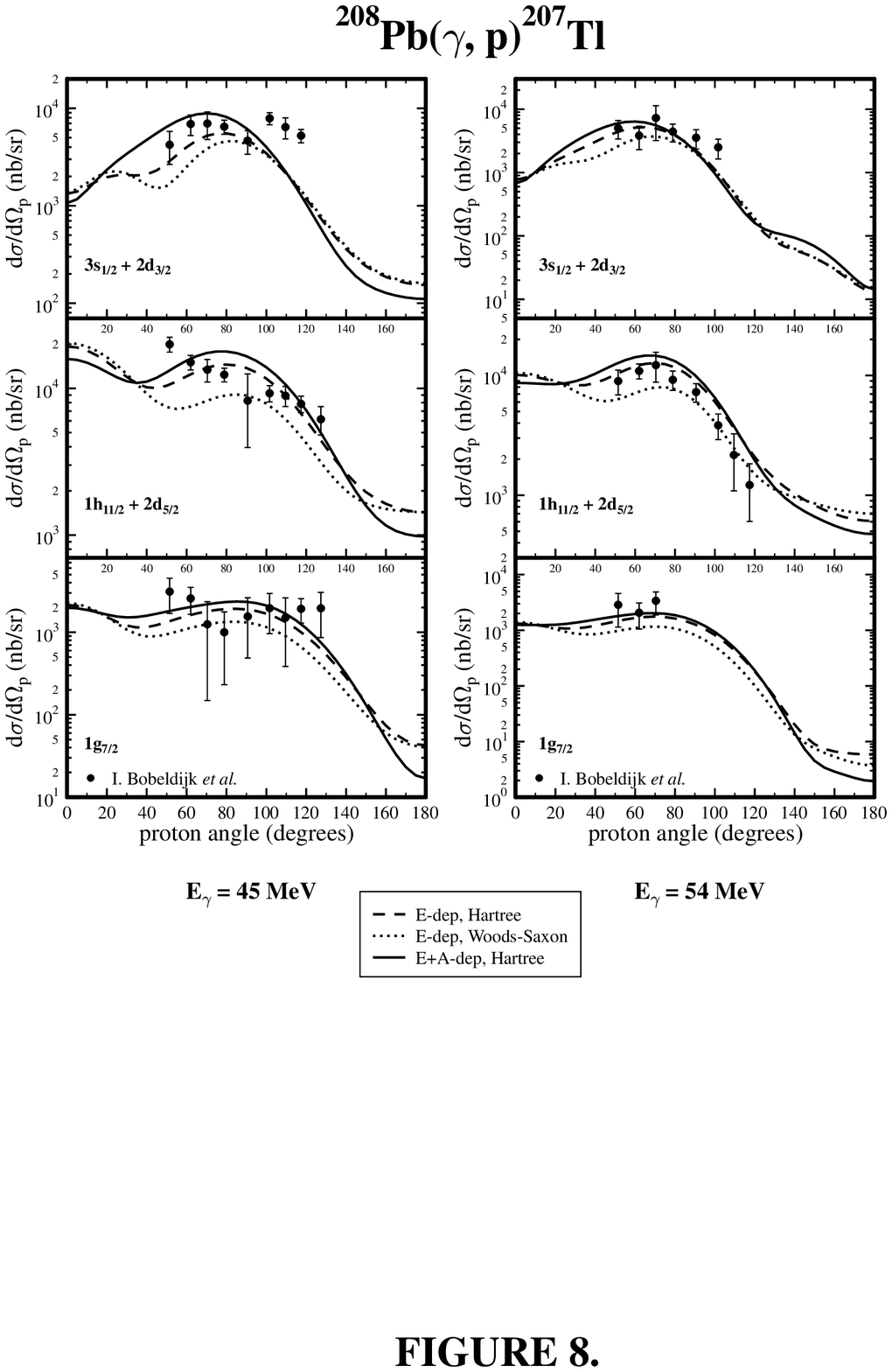}
\end{picture}
\end{figure}

\end{document}